\input amstex
\documentstyle{gen-p}
\NoBlackBoxes

\topmatter
\title Blending two discrete integrability criteria:
 singularity confinement and algebraic entropy\endtitle
\author S. Lafortune, A. Ramani, B. Grammaticos, Y. Ohta and
K.M. Tamizhmani\endauthor
\rightheadtext{BLENDING TWO DISCRETE INTEGRABILITY CRITERIA}
\leftheadtext{S. LAFORTUNE, A. RAMANI, B. GRAMMATICOS, Y. OHTA AND
K.M. TAMIZHMANI}%

\address LPTMC et GMPIB,  Universit\'e Paris VII,
 Tour 24-14, 5$^e$\'etage, 75251 Paris, France \endaddress

\curraddr CRM, Universit\'e de Montr\'eal,
Montr\'eal, C.P.6128, Succ. Centre-Ville, H3C 3J7, Canada\endcurraddr 

\email lafortus\@crm.umontreal.ca\endemail

\thanks 
S. Lafortune was supported by FCAR du Qu\'ebec for
his Ph.D. and by ``Programme de Soutien
de Cotutelle de Th\`ese de doctorat du Gouvernement du Qu\'ebec'' for his
stay in Paris.\endthanks

\address CPT, Ecole Polytechnique, CNRS, UMR 7644, 91128 Palaiseau, France
\endaddress

\email ramani\@orphee.polytechnique.fr\endemail

\address GMPIB, Universit\'e Paris VII, Tour 24-14, 5$^e$\'etage, case 7021,
 75251 Paris, France\endaddress
 
 \email grammati\@paris7.jussieu.fr\endemail
 
 \address Department of Applied Mathematics, Faculty of Engineering, 
 Hiroshima University, 1-4-1 Kagamiyama, Higashi-Hiroshima, 739-8527 
 Japan\endaddress
 
  \address Departement of Mathematics, Pondicherry University, Kalapet, 
  Pondicherry, 605014 India\endaddress

\def\yup{y_{n+1}}
\def\xup{x_{n+1}}
\def\x{x_{n}}
\def\y{y_{n}}

\def\xdo{x_{n-1}}

\def\sc{singularity confinement }
\def\zdo{z_{n-1}}
\def\zup{z_{n+1}}
\def\ado{a_{n-1}}
\def\aup{a_{n+1}}
\def\bdo{b_{n-1}}
\def\bup{b_{n+1}}

\abstract
We confront two integrability criteria for rational mappings. The first is
the singularity confinement based on the
requirement that every singularity, spontaneously appearing during the
iteration of a mapping, disappear after some steps.
The second recently proposed is the algebraic entropy criterion associated
to the growth of the degree of the iterates.
The algebraic entropy results confirm the previous findings of singularity
confinement on discrete Painlev\'e equations.
The degree-growth methods are also applied to linearisable systems. The
result is that systems integrable through
linearisation have a slower growth than systems integrable through
isospectral methods. This may provide a valuable
detector of not just integrability but also of the precise integration
method. We propose an extension of the Gambier
mapping in $N$ dimensions. Finally a dual strategy for the investigation of
the integrability of discrete systems is
proposed based on both singularity confinement and the low growth requirement.
\endabstract

\endtopmatter

\document

\head 1. Introduction \endhead

The detection of discrete integrability has for the past few years relied
on the criterion of singularity confinement [1].
This criterion is based on the observation that for integrable discrete
systems any singularity that appears
spontaneously disappears after some iterations. Let us give one example. In
the mapping
$$
\xup+\xdo={a\over \x}+{1\over \x^2},
\eqno (1.1)
$$
where $a$ is constant, we assume that, at some iteration, $\xdo$ vanishes
while $x_{n-2}$ is finite. This has as
consequence a diverging
$\x$, and iterating once more a vanishing $x_{n+1}$. If we try to compute
$x_{n+2}$ we obtain an indeterminate form,
$\infty -\infty$. In order to lift the indeterminacy we introduce a small
parameter $\epsilon$ and assume that
$\xdo=\epsilon$. Iterating we obtain successively, expanding in $\epsilon$,
$\x=1/\epsilon^2+a/\epsilon-x_{n-2}+{\Cal O}(\epsilon)$,
$\xup=-\epsilon+a\epsilon^2+{\Cal {O}}(\epsilon^3)$, $x_{n+2}=x_{n-2}+2(a
x_{n-2}+1)\epsilon+{\Cal O}(\epsilon^2)$.
By taking the appropriate limit $\epsilon\rightarrow 0$ we find that
$x_{n+2}$ is finite, and moreover contains the
information on the initial data i.e. $x_{n-2}$. The subsequent $\x$'s are
indeed finite and the singularity is confined
to the sequence $\{0,\infty^2, 0\}$, the square in the infinite sign being
a reminder of the $1/\epsilon^2$.
Singularity confinement is tailored for rational mappings although this
should not be
considered as an absolute restriction. What singularity confinement {\sl
cannot} treat are mappings which do {\sl not}
possess singularities. For example a mapping of the form
$\xup=\lambda
\x(1-\x)$ lies beyond the reach of the confinement method [2]. Our tacit
conjecture has always been
that polynomial mappings, like the logistic one, are {\sl not} integrable,
with only exception the linear one [3].

While confinement has turned out to be a most useful integrability
detector, its application was based on the unwarranted
assumption that its necessary character will be constraining enough to make
it sufficient. Still, from the outset it was
clear that confinement was not a sufficient criterion. Let us recall the
example presented already in [4]. If we examine
the singularity structure of the mapping $\xup=f(\x,n)$
where the $f$ is rational in $\x$ and analytic in $n$ we find that all
$f$'s of the form $f=\sum_k{\alpha_k \over
(\x+\beta_k)^{\nu_k}}$, with $\nu_k \in \Bbb{Z}$, lead to confinement.
Clearly not all these mappings can be integrable
and, indeed, only the homographic one, $\xup=\alpha+{\lambda\over
\x+\beta}$, is. The non sufficient character of
singularity confinement was put to a wider perspective by Hietarinta and
Viallet [5] who verified that the mapping:
$$
\xup+\xdo=\x+{1\over \x^2}
\eqno (1.2)
$$
is confining, the singularity sequence being $\{0, \infty^2, \infty^2,
0\}$, and at the same time exhibits chaotic
behaviour. Moreover this is not an isolated example and one can easily
construct mappings which are confining without
being integrable.

Clearly a more stringent integrability criterion was needed and the authors
of [5] (see also [6]), have proposed one
based on the ideas of Arnold and Veselov.
According to Arnold [7] the
complexity (in the case of mappings of the plane) is the number of
intersection points of a fixed curve with the
image of a
second curve obtained under the mapping at hand. While the complexity grows
exponentially with the iteration for
generic
mappings, it can be shown [8] to grow only polynomially for a large class
of integrable mappings. As Veselov
points out,  ``integrability has an essential correlation with the weak
growth of certain characteristics''.
Thus the authors of [5]  proposed to directly test the degree of the
successive iterates
and introduced the notion of algebraic entropy.
 It is defined as $E=\lim_{n\to \infty} \log(d_n)/n$ where $d_n$ is the
degree of the $n$th
iterate of some initial data under the action of the mapping. Since a
generic nonintegrable mapping exhibits
exponential degree-growth, a nonzero algebraic entropy indicates
nonintegrability. Integrable mappings must have zero
algebraic entropy, associated to slower-than-exponential (typically
polynomial) degree-growth. It is expected that the
requirement of zero algebraic entropy is strong enough to be a sufficient
integrability condition.

In order to make the ideas clearer we present here our method for the study
of the degree-growth. It presents a few
differences with respect to the one of [5] which, we feel, make its
practical implementation simpler. We start by
introducing homogeneous variables through the appropriate choice of initial
data. Typically, in a three-point mapping we
choose to introduce $x_0=r$, $x_1=p/q$ and thus the weight of $r$ is 0
while that of $p$, $q$ are taken
equal to 1. We then compute the homogeneous degree in $p$, $q$ in the
denominator and
numerator of $\x$ at every iteration.
Let us follow the first few iterations of mapping  (1.1) in order to
understand the mechanism of the degree-growth. We
find
$$
\displaylines{
x_2={q^2+apq-rp^2\over  p^2},\quad x_3={pP_4\over  q(q^2+apq-rp^2)^2},\cr
x_4={(q^2+apq-rp^2)P_6\over  P_4^2}, \quad x_5={P_4P_9\over qP_6^2}\cr
}$$
 where the $P_k$'s are homogeneous polynomials in $p$, $q$ of degree $k$.
(Remember that $r$ is of zero homogeneous
degree  in our
convention). A pattern becomes apparent. Whenever a new polynomial appears
in the numerator of $x_n$, its square will
appear
in the denominator of $x_{n+1}$ and it will appear one last time as a
factor of the numerator of $x_{n+2}$, after
which it
disappears due to factorisations. The singularities we are working with in
the singularity confinement approach correspond
to the zeros of  any of
these polynomials, which explains the pattern $\{0,\infty^2, 0\}$.
The singularity confinement is intimately related to this factorisation
which plays a crucial role in the algebraic
entropy approach. If we calculate the degree of the iterates, we obtain: 0,
1, 2, 5,
8, 13, 18, 25, 32, 41, \dots Clearly the degree-growth is polynomial: we have
$d_{2m}=2m^2$ and $d_{2m+1}=2m^2+2m+1$. (A remark is necessary at this
point. In order to
obtain a closed-form expression for the degrees of the iterates, we start
by computing a sufficient number of them. Once the
expression of the degree has been heuristically established we compute the
next few ones and check that they agree
with the analytical expression predicted). Thus the algebraic entropy of
this mapping is zero
in agreement with its integrable character [9].

On the other hand, if we iterate mapping  (1.2) we obtain the sequence
$$
\hskip-1.2cmx_2={p^3-p^2
qr+q^3 \over  p^2 q},\quad x_3={P_8\over  p^2(p^3-p^2qr+q^3)^2},$$
$$
 x_4={pP_{22}\over q(p^3-p^2qr+q^3)^2{P_{8}}^2},\quad
 x_5={(p^3-p^2qr+q^3)P_{58}\over q{P_{8}}^2{P_{22}}^2}.
$$
Again some factorisations and simplifications do occur which explain why
this mapping has the confinement property.
However the degree of the new terms appearing at every iteration grows too
rapidly and thus the simplifications cannot
curb the exponential growth. The degree sequence now is
$0,1,3,8,23,61,160,421,\dots$, i.e. the degrees obey the
relation $d_{n+1}-3d_n+d_{n-1}={2\over 3}(1+j^{n+1}+j^{2(n+1)})$ where $j$
is a complex cubic root of $1$ leading to an
exponential growth with asymptotic ratio $(3+\sqrt{5})/2$ and algebraic
entropy $\log((3+\sqrt{5})/2)$. The situation is
even worse in the case of nonconfining mappings where no simplifications
occur and the growth rate is maximal.

Since a considerable volume of results on integrable discrete systems were
obtained with the method of singularity
confinement it is natural to question their integrability in the light of
the findings concerning the non sufficiency
of the singularity confinement criterion. One of the aims of this review is
to provide the confirmation of these
results based on the more stringent criterion of algebraic entropy. In
particular we shall examine several
discrete Painlev\'e equations that we have obtained over the years and
obtain their growth properties. The main result of
this analysis is that, in every case studied, the singularity confinement
criterion {\sl when used for deautonomisation of
an integrable mapping}, turns out to be sufficient [10]. This means that an
equation, the autonomous form of which is
integrable and which has been deautonomised following the constraints
provided by singularity confinement, leads to
exactly the same degree-growth as the initial, autonomous, one. Thus we
expect the deautonomised  equation to be as
integrable as its autonomous limit. This finding is of capital importance
since it validates the results previously
obtained with the singularity confinement approach. Moreover it sets the
frame for a new, composite, approach.
Whenever we wish to obtain the integrable cases of an equation containing a
certain number of parameters we can perform
a first, exploratory, study using the confinement method. Once the
confinement constraints have been used in order to
limit the freedom, the algebraic entropy, or low-growth criterion, can be
implemented in order to pin down the truly
integrable cases. Examples of this dual approach can be found in [11].

\head 2. Discrete Painlev\'e equations\endhead

Let us first recall what has always been our approach to the derivation of
discrete Painlev\'e equations. We
start from an autonomous system the integrability of which has been
independently established. In the case of discrete
Painlev\'e equations, this system is the QRT mapping [9]:
$$ f^{(1)}(x_n)-(\xup+\xdo)f^{(2)}(x_n)+\xup\xdo f^{(3)}(x_n)=0\eqno(2.1)$$
When the $f^{(i)}$'s are quartic functions, satisfying specific
constraints, the mapping (2.1) is integrable in terms of elliptic
functions. Since the elliptic functions are the
autonomous limits of the Painlev\'e transcendents, the mapping (2.1) is the
appropriate starting point for the construction
of the nonautonomous discrete systems which are the analogues of the
Painlev\'e equations. The procedure we used, often
referred to as `deautonomisation', consists in finding the dependence of
the coefficients of the quartic polynomials
appearing in (2.1) with respect to the independent variable $n$, which is
compatible with the \sc property. Namely, the
$n$-dependence is obtained by asking that the singularities are indeed
confined. The reason why this procedure can be
justified is the following. Since the autonomous starting point is
integrable, it is expected that the growth of the degree of the
iterates is polynomial. Now it turns out that the application of the \sc
deautonomisation corresponds to the requirement that
the nonautonomous mappings lead to the same factorizations and subsequent
simplifications and have precisely the same growth
properties as the autonomous ones.  These considerations will be made more
transparent thanks to the examples we present in
what follows.

Let us start with a simple case. We consider the mapping:
$$\xup+\xdo={a\over x_n}+{1\over \x^2} \eqno(2.2)$$
where $a$ depends on $n$.
The \sc result is that $a$ must satisfy the conditions
 $\aup-2a_n+\ado=0$ i.e. $a$ is linear in $n$.
Assuming now that $a$ is an arbitrary function of $n$ we compute the
iterates of (2.2). We obtain the sequence:
$$
\displaylines{
x_2={q^2+a_1pq-p^2r\over  p^2},\quad x_3={pQ_4\over
q(q^2+a_1pq-p^2r)^2},\cr\hfill
x_4={(q^2+a_1pq-p^2r)Q_7\over pQ_4^2}, \quad
 x_5={pQ_4Q_{12}\over q(q^2+a_1pq-p^2r)Q_7^2}
}$$
where the $Q_k$'s are homogeneous polynomials in $q$, $r$ of degree $k$.
The simplifications that do occur are
insufficient to curb the asymptotic growth. As a matter of fact, if we
follow a particular factor we can check that it keeps
appearing either in the numerator or the denominator (where its degree is
alternatively 1 and 2). This corresponds to the
unconfined singularity pattern
$\{0,\infty^2,0,\infty,0,\infty^2,0,\infty,\dots\}$. The confinement
condition
$\aup-2a_n+\ado=0$ is the condition for $q$ to divide exactly $Q_7$, for
both $q$ and $r^2+a_1qr-pq^2$ to divide exactly
$Q_{12}$, etc.
 Let us now turn
to the computation of the degrees of $\x$. We obtain
successively 0, 1, 2, 5, 10, 21, 42, 85,\dots. The growth is exponential,
the degrees behaving like
$d_{2m-1}=(2^{2m}-1)/3$  and $d_{2m}=2d_{2m-1}$, a clear indication that
the mapping is not integrable in
general. Already at the fourth iteration the  degrees differ in the
autonomous and nonautonomous cases.  Our approach consists in requiring
that the degree in the nonautonomous case be {\sl
identical} to the one obtained in the autonomous one. If we implement the
requirement that $d_4$ be 8 instead of 10 we
find the condition
$\aup-2a_n+\ado=0$, i.e. precisely the one obtained through singularity
confinement. Moreover, once this condition
is satisfied, the subsequent degrees of the nonautonomous case coincide
with that of the autonomous one. Thus this
mapping, leading to polynomial growth, should be integrable, and, in fact,
it is. As we have shown in [9], where we
presented its Lax pair, equation (2.2) with $a_n=\alpha n+\beta$ is a
discrete form of the Painlev\'e I equation. In the
examples that follow, we shall show that in all cases the nonautonomous
form of an integrable mapping obtained through
\sc leads to exactly the same degrees of the iterates as the autonomous one.

We now turn to what is known as the ``standard'' discrete Painlev\'e
equations [12] and compare the results of \sc to
those of the algebraic entropy approach. We start with d-P$_{\text I}$ in the
form:
$$\xup+\x+\xdo=a+{b\over \x}. \eqno(2.3)$$
The degrees of the iterates of the autonomous mapping are 0, 1, 2, 3, 6, 9,
12, 17, 22, \dots, i.e. a quadratic growth with
$d_{3m+k}$=$3m^2+(2m+1)k$, for $k=$0,1,2, while those of the
generic  nonautonomous one are 0, 1, 2, 3, 6, 11, \dots Requiring two extra
factorisations at that level (so as to bring
$d_5$ down to 9) we find the following conditions $\aup=a_n$, so $a_n$ must
be a constant, and $b_{n+2}-\bup-b_n+\bdo$=0,
i.e. $b_n$ is of the form $b_n=\alpha n+\beta+\gamma (-1)^n$ which are
exactly the result of singularity confinement.
Implementing these conditions we find that the autonomous and nonautonomous
mappings have the same (polynomial) growth
[10]. Both are integrable, the Lax pair of the nonautonomous one, namely
d-P$_{\text I}$ having been given in [13,14,15].

For the discrete P$_{\text II}$ equation we have
$$\xup+\xdo={ax_n+b\over x_n^2-1} \eqno(2.4)$$
The degrees of the iterates in the autonomous case are $d_n$=0, 1, 2, 4, 6,
9, 12, 16, 20, \dots, (i.e. $d_{2m-1}$=$m^2$,
$d_{2m}$=$m^2+m$) while in the generic nonautonomous case we find the first
discrepancy for $d_4$ which is now 8. To
bring it down to 6 we find two conditions, $\aup-2a_n+\ado$=0 and
$\bup=\bdo$. This means that $a$ is linear in $n$ and $b$ is
an even/odd constant, as predicted by singularity confinement. Once we
implement these constraints, the degrees of the
nonautonomous and autonomous cases coincide. The Lax pair of equation (2.4)
in the nonautonomous form, i.e.
d-P$_{\text II}$, has been presented in [13,16,17].

The $q$-P$_{\text III}$ equation was obtained from the deautonomisation of
the mapping:
$$\xup\xdo={(x_n-a)(x_n-b)\over (1-cx_n)(1-x_n/c)} \eqno(2.5)$$
In the autonomous case we obtain the degrees $d_n$=0, 1, 2, 5, 8, 13, 18,
\dots, just like for equation (2.2), while in  the
generic nonautonomous case we have 0, 1, 2, 5, 12,\dots  For $d_4$ to be 8
instead of 12, one needs four factors to
cancel out. The conditions are $c_{n+1}=c_{n-1}$ and
$\aup\bdo=\ado\bup=a_nb_n$. Thus $c$ is a constant up to an  even/odd
dependence, while $a$ and $b$ are proportional to $\lambda^n$ for some
$\lambda$, with an extra even/odd dependence, just as
predicted by
\sc in [12]. The Lax pair for $q$-P$_{\text III}$ has been presented in
[13,18].

For the remaining three discrete Painlev\'e  equations the Lax pairs are
not known yet. It is thus important to have one more
check of their integrability provided by the algebraic entropy approach. We
start with d-P$_{\text IV}$ in the form:
$$(\xup+x_n)(\xdo+x_n)={(x_n^2-a^2)(x_n^2-b^2)\over (x_n+z_n)^2-c^2}
\eqno(2.6)$$
where $a$, $b$ and $c$ are constants. If $z_n$ is constant we obtain for
the degrees of the successive iterates $d_n$=0, 1, 3, 6,
11, 17, 24, \dots The general expression of the growth is $d_n$=6$m^2$ if
$n=3m$, $d_n$=6$m^2+4m+1$ if $n=3m+1$ and
$d_n$=6$m^2+8m+3$ if $n=3m+2$.  This polynomial (quadratic) growth is
expected since in the autonomous case this equation is
integrable, its solution being given in terms of elliptic functions. For a
generic
$z_n$ we obtain the sequence  $d_n$=0, 1, 3, 6, 13,
\dots The condition for the extra factorizations to occur in the last case,
bringing down the degree $d_4$ to 11, is for
$z$ to be linear in $n$. We can check that the subsequent degrees coincide
with those of the autonomous case.

For the discrete Painlev\'e V equation we start from:
$$(\xup x_n-1)(\xdo x_n-1)={(x_n^2+ax_n+1)(x_n^2+bx_n+1)\over
(1-z_ncx_n)(1-z_ndx_n)} \eqno(2.7)$$
where $a$, $b$, $c$ and $d$ are constants. If moreover $z$ is also a
constant, we obtain exactly the same sequence  of degrees
$d_n$=0, 1, 3, 6, 11, 17, 24, \dots, as in the d-P$_{\text IV}$ case.  Again,
this polynomial (quadratic) growth is expected
since this mapping is also integrable in terms of elliptic functions. For
the generic nonautonomous case we again find the
sequence $d_n$=0, 1, 3, 6, 13, \dots Once more we require a factorization
bringing down $d_4$ to 11. It turns out that
this entails a $z$ which is exponential in $n$, which then generates the
same sequence of degrees as the autonomous case.
In both the d-P$_{\text IV}$ and $q$-P$_{\text V}$ cases we find the
$n$-dependence already obtained through singularity
confinement. Since this results to a vanishing algebraic entropy we expect
both equations to be integrable.

The final system we shall study is the one related to the discrete P$_{\text
VI}$ equation:
$${(\xup x_n-\zup z_n)(\xdo x_n-\zdo z_n)\over (\xup x_n-1)(\xdo
x_n-1)}={(x_n^2+az_nx_n+z_n^2)(x_n^2+bz_nx_n+z_n^2)\over
(x_n^2+cx_n+1)(x_n^2+dx_n+1)}\eqno(2.8)$$ where $a$, $b$, $c$ and $d$ are
constants. In fact the generic symmetric
QRT mapping can be brought to the  autonomous ($z_n$ constant) form of
equation (2.8) through the appropriate homographic
transformation.  In the autonomous case, we obtain the degree sequence
$d_n$=0, 1, 4, 9, 16, 25, \dots, i.e. $d_n$=$n^2$. Since mapping (2.8) is
rather complicated we cannot investigate its full
freedom. Still we were able to perform two interesting calculations. First,
assume that  in the rhs instead of the function $z_n$
a different function $\zeta_n$ appears. In this case the degrees grow like
0, 1, 5, \dots, and the condition to have $d_2$=4
instead of 5 is
$\zup\zdo z_n^2=\zeta_n^4$. Assuming this is true, we compute the degree
$d_3$ of the next iterate and find $d_3$=13 instead of 9.
To bring down $d_3$ to the value 9 we need $z_n^2=\zeta_n^2$, which up to a
redefinition of $a$ and $b$ means $z_n=\zeta_n$. This
implies $\zup\zdo=z_n^2$, and $z_n$ is thus an exponential function of $n$,
$z_n$=$z_0\lambda^n$ (which is in agreement
with the results of [19]). Then  a quartic factor drops out and $d_3$ is
just 9. One can then check that the next degree
is 16, just as in the autonomous case. Thus the $q$-P$_{\text VI}$ equation
leads to the same growth as the generic
symmetric QRT mapping and is thus expected to be integrable.
As a matter of fact we were able to show that the generic {\sl asymmetric}
QRT mapping leads to the same growth $d_n$=$n^2$ as
the symmetric one. This is not surprising, given the integrability of this
mapping.
What is interesting is that the growth of the generic  symmetric and
asymmetric QRT mappings are the same. Thus $d_n$=$n^2$ is
the maximal growth one can obtain for the QRT mapping in the homogeneous
variables we are using. As a matter of fact we have
also checked that the asymmetric nonautonomous $q$-P$_{\text VI}$ equation,
introduced in [19] led to exactly the same
degree-growth $d_n$=$n^2$.

The \sc results have been confronted to the algebraic entropy approach for
several other discrete Painlev\'e
equations. In every case examined the deautonomisation obtained has turned
out to be the right one, it was the
condition for the degree-growth to be identical to the one of the
autonomous system. Thus despite the non sufficiency of
the
\sc the integrability predictions for discrete Painlev\'e equations based on
this criterion are confirmed.

\head 3. Linearisable equations\endhead

In this section we shall examine this particular class of mappings which
are linearisable and study their growth
properties. Most of these systems were obtained using the singularity
confinement criterion and thus a study of the growth
of the degree of the iterates would be an interesting complementary
information. Moreover, as we will show, the
linearisable systems do possess particular growth properties which set them
apart from the other integrable discrete
systems.

The first mapping we are going to treat is a two-point mapping of the form
$x_{n+1}=f(x_n,n)$ where
$f$ is rational in $x_n$ and analytical in $n$. As explained in the
introduction for all $f$'s of the form
$\sum_k {\alpha_k\over (x_n+\beta_k)^{\nu_k}}$ the singularity confinement
requirement is satisfied. However only the
discrete Riccati, $\xup=\alpha+{\lambda\over \x+\beta}$, is expected to be
integrable. Our argument in [20], for
the rejection of these confining but nonintegrable cases, was based on the
proliferation of the preimages of a given
point. If we solve the  mapping for $x_n$ in terms of $x_{n+1}$ we do not
find a uniquely defined $x_n$ and, iterating,
the number of $x_{n-k}$ grows exponentially. In what follows we shall
analyse this two-point mapping in the  light of the
algebraic entropy approach. We start from the simplest case which we expect
to be nonintegrable,
$$
\xup=\alpha+{\lambda\over \x+\beta}+{\mu\over \x+\gamma}.
\eqno(3.1)
$$ The initial condition we are going to iterate is
$x_0=p/q$ and the degree we calculate is the homogeneous  degree in $p$ and
$q$ of the numerator (or the denominator) of the iterate. We obtain readily
the following degree sequence
$d_n=1,2,4,8,16,\dots$ i.e. $d_n=2^n$. Thus the algebraic entropy of the
mapping is $\log (2)>0$, an indication that the
mapping cannot be integrable. Now we ask how can one curb the growth and
make it
nonexponential. It turns out that the only possibilities are
$\lambda\mu=0$ or
$\beta=\gamma$. In either case mapping (3.1) becomes a homography. The
degree in this case is simply $d_n=1$ for all $n$.
This is an interesting result, clearly due to the fact that the homographic
mapping is linearisable through a simple
Cole-Hopf transformation.

The second mapping we shall examine is one due to Bellon and collaborators [21]
$$
\eqalign{
\xup={\x+\y-2\x\y^2 \over \y(\x-\y)}, \cr
\yup={\x+\y-2\x^2\y \over \x(\y-\x)}. }
\eqno(3.2)
$$ The degree-growth in this case is studied starting from $x_0=r$,
$y_0=p/q$ and again we calculate the homogeneous
degree of the iterate in $p$ and $q$, i.e. we set the degree of $r$ to
zero. (Other choices could have been possible but
the conclusion would not depend on these details.) We obtain the degrees
$d_{x_n}=0,2,2,4,4,6,6,\dots$ and
$d_{y_n}=1,1,3,3,5,5,\dots$ i.e. a linear degree-growth. This is in perfect
agreement with the integrable character of the
mapping. As was shown in [22] it does satisfy the unique preimage
requirement and possesses a constant of motion
$k={1-\x\y\over \y-\x}$, the use of which reduces it to a homographic
mapping for $\x$ or $\y$.

The third mapping we are going to study is the one proposed in [20]
$$
\eqalign{
\xup={\x(\x-\y-a)\over \x^2-\y}, \cr
\yup={(\x-\y)(\x-\y-a) \over \x^2-\y}}\eqno(3.3)
$$ where $a$ was taken constant. We start by assuming that $a$ is an
arbitrary function of $n$ and compute the growth of
the degree. We find $d_{x_n}=0,1,2,3,4,5,6,7,8,\dots$ and
$d_{y_n}=1,2,3,4,5,6,7,8,9,\dots$ i.e. again a linear growth. This is an
indication that (3.3) is integrable for
arbitrary
$a_n$ and indeed it is. Dividing the two equations we obtain
${\yup/\xup}=1-{\y/\x}$ i.e. ${\y/\x}=1/2+k(-1)^n$
whereupon (3.3) is reduced to a homographic mapping for $x$. Thus in this
case the degree-growth has successfully
predicted integrability.

A picture starts emerging at this point. While in our study of discrete
Painlev\'e equations and the QRT mapping we found
quadratic growth of the degree of the iterate, linearisable second-order
mappings seem to lead to slower growth. In order
to investigate this property in detail we shall analyse the three-point
mapping we have studied in [4,23] from the point
of view of integrability in general and linearisability in particular. The
generic mapping studied in [23] was one
trilinear in $\x$,
$\xup$, $\xdo$. Several cases were considered. Our starting point is the
mapping,
$$
\xup \x \xdo +\beta \x\xup+\zeta\eta\xup\xdo+\gamma \x\xdo+\beta\gamma
\x+\eta \xdo+\zeta \xup+1=0.
\eqno(3.4)
$$
We start with the initial conditions $x_0=r$, $x_1=p/q$ and compute the
homogeneous degree in $p$, $q$ at every $n$.
We find
$d_n=0,1,1,2,3,5,8,13,\dots$ i.e. a Fibonacci sequence
$d_{n+1}=d_n+d_{n-1}$ leading to exponential growth of $d_n$ with
asymptotic ratio ${1+\sqrt{5}\over 2}$. Thus mapping (3.4) is not expected
to be integrable in general. However, as shown
in [23] integrable subcases do exist. We start by requiring that the
degree-growth be less rapid and as a drastic decrease
in the degree we demand that $d_3=1$ instead of $2$. We find that this is
possible when either $\beta=\zeta=0$ in which
case the mapping reduces to:
$$
\xup=-\gamma-{\eta \over \x}-{1\over \x\xdo}
\eqno(3.5)
$$
or $\gamma=\eta=0$, giving a mapping identical to (3.5) after $x\rightarrow
1/x$. In this case the degree is $d_n=1$ for
$n>0$. Equation (3.5) is the generic projective three-point mapping,
written in canonical form. Its linearisation can be
obtained [23] in terms of a system of three linear equations, a fact which
explains the
constancy of the degree.

Non generic subcases of (3.4) some of which are integrable do exist. They
have been studied in [23,24].

Linearisability through the reduction to a projective system is not the
only possibility. Other possibilities do exist.
Let us start by considering  the generic three-point mapping that can be
considered as the discrete derivative of a
(discrete) Riccati equation. Let us start from the general homographic
mapping which we can write as
$$
A\x\xup+B\x+C\xup+D=0
\eqno(3.6)
$$
where $A,B,C,D$ are linear in some constant quantity $\kappa$.
In order to take the discrete derivative we extract the constant $\kappa$
and rewrite  (3.6) as:
$$
\kappa={\alpha\x\xup+\beta\x+\gamma\xup+\delta\over\epsilon\x\xup+\zeta\x+\eta\x
up+\theta}.
\eqno (3.7)
$$
Using the fact that $\kappa$ is a constant, it is now easy to obtain the
discrete derivative by downshifting  (3.7)
and subtracting it form  (3.7) above.
Instead of examining this most general case we concentrate on the forms
proposed in [25].
They correspond to the reduction of  (3.7) to the two cases:
$$\kappa=\xup+a+{b\over\x}\eqno (3.8)$$
$$\kappa={\xup(\x+a)\over\x+b}\eqno (3.9)$$
Next we compute the discrete derivatives of  (3.8) and  (3.9). We find:
$$\xup=\x+a_{n-1}-a_n-{b_n \over \x}+{b_{n-1}\over \xdo}\eqno (3.10)$$
and
$$\xup=\x{\xdo+a_{n-1}\over\x+a_n}{\x+b_n \over \xdo+b_{n-1}}.\eqno (3.11)$$

The study of the degree of growth of  (3.10) and  (3.11) can be performed
in a straightforward way. For both mappings we
find the sequence
 $d_n=0,1,2,3,4,5,6,\dots$ i.e. a linear growth just as in the cases of
mappings (3.2) and (3.3). If we substitute
${b_{n-1}}$ by
${c_{n-1}}$ in the last term of the rhs of  (3.10) or the denominator of
(3.11) we find $d_n=0,1,2,4,8,16,\dots$ i.e.
$d_n=2^n$ for $n>0$ unless
$c=b$. Investigating all the possible ways to curb the growth we find for
both  (3.10) and  (3.11) that $c=0$ is
also a possibility to bring $d_3$ down to 3. However a detailed analysis
of this case shows that for $c=0$ we have
$d_n=0,1,2,3,5,8,13,21,\dots$ i.e. a Fibonacci sequence with slower, but
still exponential, growth (i.e. ratio
${1+\sqrt{5}\over 2}$ instead of $2$).
\medskip
\noindent {\it A bonus study: the $N\!$-dimensional Gambier mapping}
\medskip
While most of the results included in this paper were presented in previous
publications of ours, in this section we
shall present a study which appear for the first time in these proceedings.
It concerns the extension of the Gambier
mapping to $N$ dimensions.

The two-dimensional Gambier mapping was introduced in [26,27] as a
discretisation of the second order differential
equation discovered by Gambier, in his study of second order ODE's having
the Painlev\'e property. The Gambier
equation is in fact a system of two Riccati's in cascade. The latter means
that the system consists in one Riccati for one
variable and a second one (for the second variable) with coefficients
depending (linearly) on the variable of the first
one. The discretisation of the Gambier system is straightforward: just
substitute homographic mapping, i.e. discrete
Riccati, in the previous sentence. Thus the general second-order Gambier
mapping is given by:
$$
y_{n+1}={a\y +b\over c\y+d},\eqno(3.12a)
$$
$$
x_{n+1}={(e\y+f)\x +(g\y+h)\over
(j\y+k)\x+(l\y+m)}
\eqno(3.12b)
$$
where $a$, $b$, $\dots$, $m$ are functions of $n$.
In [27] we have studied this mapping in detail from the point of view of
the singularity structure. This was motivated by
the fact that we aimed at being able to express the solution as an infinite
product of matrices, {\sl even across
singularities}. On the other hand if we are not interested in this fine
point, the linearisability of (3.12) can be
obtained through a Cole-Hopf transformation for each variable.

The study of the degree-growth of (3.12) is straightforward. We start from
$x_0=r$, $y_0=p/q$ and compute the homogeneous
in $p,q$ degree of (4.1a) and (4.1b). Since (3.12a) is a Riccati its degree
does not grow i.e. we have $d_{y_n}=1$. Given
the structure of (3.12b) we have $d_{x_{n+1}}=d_{x_n}+d_{y_n}$ and thus
$d_{x_n}=n$. What is interesting here is that the
Gambier mapping exhibits a linear degree-growth independently of the
precise values of $a$, $b$, $\dots$, $m$. The fact
that it can be reduced to Riccati's in cascade is enough to guarantee its
integrability.

The generalisation of (3.12) to $N+1$ dimensions is straightforward. We
find the system:
$$
\eqalign{
\hskip-2.5cm x^{(0)}_{n+1}&={a^{(0)}\x^{(0)} +b^{(0)}\over
c^{(0)}\x^{(0)}+d^{(0)}},\cr
x^{(i)}_{n+1}&={(e^{(i)}\x^{(i-1)}+f^{(i)})\x^{(i)}
+(g^{(i)}\x^{(i-1)}+h^{(i)})\over
(j^{(i)}\x^{(i-1)}+k^{(i)})\x^{(i)}+(l^{(i)}\x^{(i-1)}+m^{(i)})},\quad i=1,
\dots, N.} \eqno(3.13)
$$

The study of the degree-growth of system (3.13) can be performed along the
$N=1$ case. From (3.13) we have the recurrence:
$
dx^{(k)}_{n+1}=dx^{(k)}_{n}+dx^{(k-1)}_{n}$. We obtain formally
$dx^{(k)}_{n+1}=\sum^{n}_{p=0}dx^{(k-1)}_p$. Thus
starting from $dx^{(0)}_n=1$ we find $dx^{(N)}_n\propto n^N$. Thus the
$(N+1)$-dimensional Gambier mapping has polynomial
growth for every dimension. The linearisation is obtained through a
Cole-Hopf transformation.

Given the structure of (3.13) it is clear that we can solve successively
for each variable and express finally (3.13) as
a single $(N+2)$-point mapping. This leads us to another integrable
discretisation of linearisable mappings with cascade
structure. Let us consider the second order system
$$
\eqalign{
 y_{n+1}&={a\y +b\over c\y+d},\cr
x_{n+1}&={f_1(\y)\x -f_2(\y)\over
f_4(\y)-f_3(\y)\x}} \eqno(3.14)
$$
where the $f_i$'s are polynomial in $\y$.
From the structure of (3.14) it is clear that it can be linearised,
independently of the precise form of the $f_i$.
Similarly the degree-growth of $\x$ is always linear. On the other hand it
is not, in general, possible to express (3.14)
as a single three-point mapping. Extension of (3.14) to $N$-dimensions can
be obtained along the lines of the
$N$-dimensional discrete Gambier system.

In this section we have studied the growth properties of various
linearisable systems identified through the \sc
criterion. In every case the \sc results were confirmed. Moreover it turned
out that the linearisable systems lead to
slower growth than systems which are integrable by other methods. This
property could be used for the classification of
integrable systems and be a valuable indication as to the precise method of
their integration.

\head 4. Conclusion\endhead

In this work we have presented a comparative review of results obtained
with the singularity confinement and the
algebraic entropy methods. While the former approach is not a sufficient
criterion of integrability, the confirmation of
its results (through the second, more stringent, criterion) in the domain
of discrete Painlev\'e equations leads to
interesting new insights. In every case examined, the singularity
confinement, when used for the deautonomisation of an
integrable autonomous mapping turned out to give sufficient constraints for
the degree-growth to be non-exponential. Thus
we can propose a new strategy for the detection of integrable discrete
systems. Given a mapping which contains several
parameters the \sc necessary criterion can be used in order to screen  it
for possible integrable discrete cases. Once
the research domain is reduced the growth properties can be studied leading
to better integrability candidates.

Another interesting result of our studies concerns the degree-growth of
linearisable systems. We found that, while for
second-order mappings the generic integrable case is associated to
quadratic growth, the linearisable mappings lead to
zero or linear growth. The growth exponent is of course a property which
depends on the dimension. We surmise that the
generic integrable $N$th order mapping will lead to growth $n^N$, while the
linearisable mappings of the same dimension
will lead to slower growth. Our study of the $N$th order Gambier system
shows that the growth is $n^{N-1}$, and we expect
the the projective system of order $N$ to lead to zero growth. Thus the
detailed study of the growth properties can
become a precious indication as to the precise method of integration of a
given discrete system.
 
\head Acknowledgements\endhead
 
  The authors are grateful to several anonymous referees who, by
insisting that the \sc results are wrong in
the light of the Hietarinta and Viallet findings, pushed us to learn the
techniques of degree-growth which allowed us
to {\sl confirm the validity of our previous results}.

\refstyle{A}

\enddocument